\documentclass[]{chjaa}                  

\usepackage{graphicx,times}             
\input{epsf.sty}                        
\input{psfig.sty}                       

\begin{document}

   \title{The World Space Observatory (WSO-UV)}

   \subtitle{Current status}

   \volnopage{Vol.0 (200x) No.0, 000--000}      
   \setcounter{page}{1}           

   \author{Michela Uslenghi
      \inst{1} \mailto{}
   \and Isabella Pagano
        \inst{2}
   \and Cristian Pontoni
        \inst{2}
   \and Salvatore Scuderi
        \inst{2}
   \and Boris Shustov
        \inst{3}
      }

   \institute{INAF -- IASF - Milano, Via E.Bassini 15, I-20133 Milano, Italy\\
             \email{uslenghi@iasf-milano.inaf.it}
        \and
             INAF -- Catania Astrophysical Observatory, Via S.Sofia 85, I-95123 Catania, Italy \\
        \and
             INASAN, Moscow, Russia \\
             }

   \date{Received~~2001 month day; accepted~~2001~~month day}

   \abstract{
This paper reports on the current status of the World Space
Observatory WSO-UV, a space mission for UV astronomy, planned for
launch at the beginning of next decade. It is based on a 1.7 m telescope, with
focal plane instruments including high resolution spectrographs,
long slit low resolution spectrographs and imaging cameras.
   }

   \authorrunning{M. Uslenghi et al. }            
   \titlerunning{The World Space Observatory}  

   \maketitle

%
%
\section{Introduction}           
\label{sect:intro}

UV
spectroscopic and imaging capabilities are fundamental for astrophysics since thermal
phenomena at temperatures T$>$10,000 K, with flux emission mostly in
the UV, occur in a wide range of astrophysical events.
Also, the electronic transitions of the most abundant molecules in
the Universe (H$_{2}$, CO, OH, CS, CO$_{2}^{+}$, CO$_{2}$) are in
the UV range. This results in UV providing the most sensitive tools
to trace the distribution of (baryonic) matter in the Universe,
other than to diagnose the chemical composition, physical properties
and kinematics of astronomical objects of all types.

In recent years, there have been three major instruments working in
the UV: \emph{Hubble Space telescope} (HST), \emph{Far Ultraviolet
Spectroscopic Explorer} (FUSE) and \emph{Galaxy Evolution Explorer}
(GALEX). The first two are observatory--like missions, whereas GALEX
is dedicated to an all--sky survey and is currently providing wide
field and low resolution spectra of a large number of astronomical
objects which will require detailed UV follow-ups. However, after
the failure of the HST STIS spectrograph in 2004, no facilities to
get medium to high resolution spectra in the classical UV domain
(1000--3000 {\AA}) have been available to the community.

Access to UV is becoming problematic. Even with the planned upgrade
of HST with the \emph{Cosmic Origin Spectrograph} (COS), during the
service mission SM4, HST is planned to work till 2013, thus posing
the problem of developing new facilities for UV astrophysics in the
post-HST era, before the advent of future large (8m class) UV
telescopes, currently under discussion but not yet included in the
plans of any space agency (and then to be scheduled unlikely before
2020--2025).

The World Space Observatory Ultraviolet -- WSO-UV -- is a multi-national
project grown out of the needs of the astronomical community to have
future access to the ultraviolet range. Planned to operate for 5(+5) years starting from 2011, it
will fill in the gap between HST and the future large UV
telescopes, complementing the wavelength coverage of the IR
\emph{James Webb Space Telescope} (JWST), which will be operative in
the same period.

\begin{figure}
   \vspace{2mm}
   \begin{center}
   \hspace{3mm}\psfig{figure=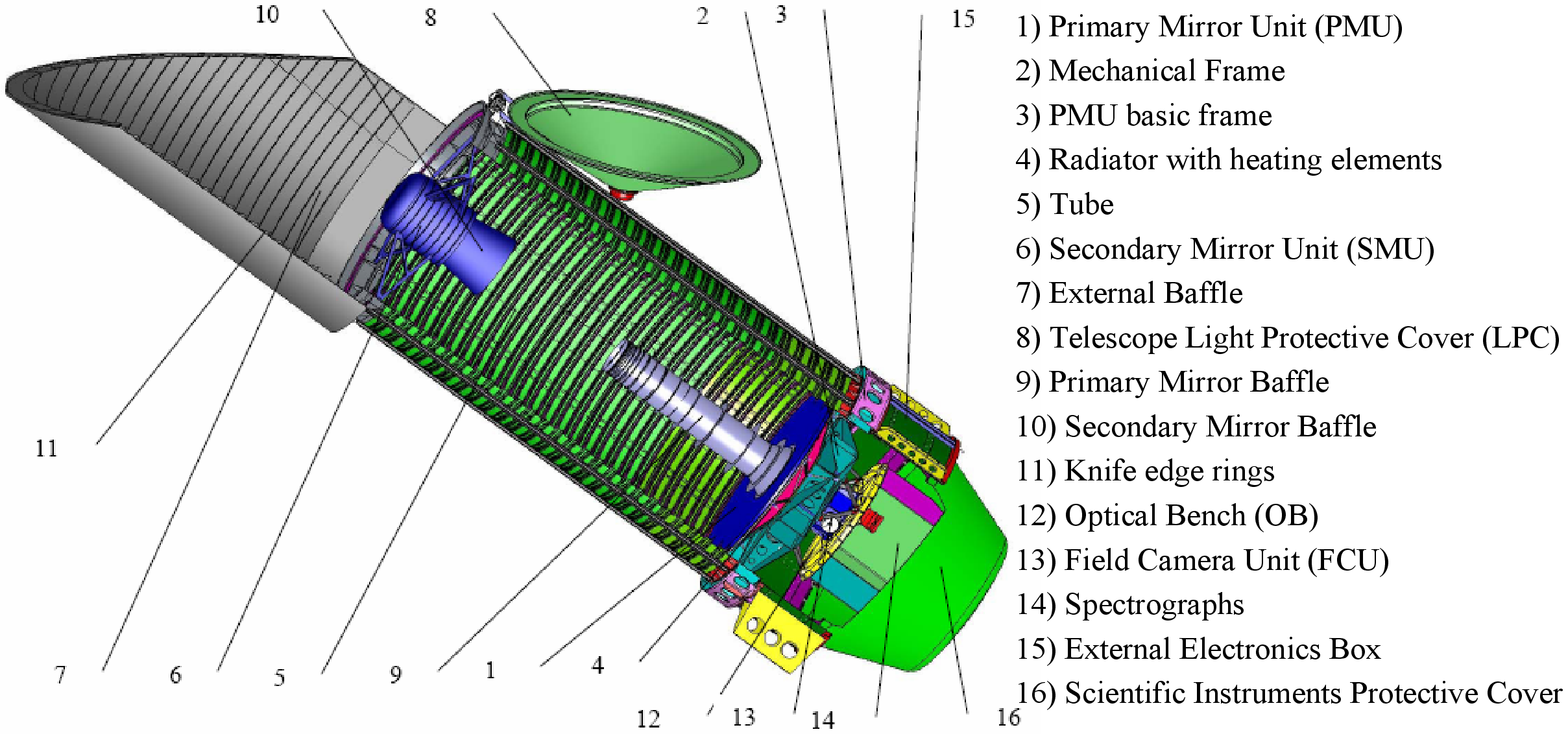,width=150mm,angle=270.0}
   \caption{General view of the T--170M telescope.   }
   \label{Fig:T170M}
   \end{center}
\end{figure}


\section{Mission Overview}
\label{sect:Mission}

\begin{table}[]
  \caption[]{ Main characteristics of WSO-UV.  }
  \label{Tab:main_char}
  \begin{center}\begin{tabular}{lr}
  \hline\noalign{\smallskip}
   \hline\noalign{\smallskip}
\textbf{Spacecraft}                   \\
  \hline\noalign{\smallskip}
Spacecraft mass with propellant  & 2900 kg \\  
Payload mass & 1600 kg \\
Instrumentation Compartment power consumption & 750 W \\
Data transmission rate (S-band) & 2 Mb/s \\
Service telemetry data transmission rate & 32 kb/s \\
Platform star tracker pointing accuracy & 30 arcsec \\
Stabilization and pointing accuracy & 0.03 arcsec \\
Spacecraft angular rate in stabilization mode & 2x10-5 degree/sec \\
Spacecraft slew rate & 0.1 degree/sec \\
Maximum duration of scientific observation in continuous mode & 30
hr \\

  \hline\noalign{\smallskip}
  \textbf{Telescope} \\
  \hline\noalign{\smallskip}
Optical System & Ritchey-Chr\'{e}tien \\
Telescope entrance pupil diameter & 1.7 m \\
Effective focal length & 17 m \\
Parameter Value F/ratio & 10 \\
FoV diameter  & 30 arcmin (148,48 mm) \\
Scale & 12,13 arcsec/mm \\
Wavelength range & 100--310 nm (with extension to the visible) \\
Primary wavelength & 200 nm\\
  \noalign{\smallskip}\hline
  \end{tabular}\end{center}
\end{table}
The World Space Observatory-UV is an international collaboration led
by Russia to build a space telescope optimized in the UV range and
devoted to investigate numerous astrophysical phenomena from
planetary science to cosmology (Barstow et al. \cite{Bar03}, Pagano
et al. \cite{Pag07}). The satellite will be based on the "Navigator"
platform, a service module used also for other Russian projects
(e.g. Elektro and RadioAstron). Table~\ref{Tab:main_char} summarizes
WSO-UV spacecraft characteristics. Telescope, launcher (Zenith 2SB)
and platform (Navigator) will be developed in Russia, whereas focal
plane instruments will be provided by Germany, China and Italy, with
contributions from UK, and Spain. Ground Segment is under design
mainly in Spain and Russia, with possible contributions from other
countries. The spacecraft will be put in a geosynchronous orbit at a
height of 35\,800 km and an inclination of 51.4 degrees.

\section{Telescope}
\label{sect:telescope}

The WSO-UV telescope T--170M is a new version of the T--170
telescope designed by Lavochkin Association (Russia) for the
Spectrum-UV mission. It is a Ritchey-Chr\'{e}tien with a 1.7m
hyperbolic primary mirror, focal ratio F/10 (platescale 12.13
arcsec/mm) and a corrected field of view of 0.5 degrees. The optical
quality of the two mirrors is $\lambda/30$ rms at 633 nm. The
primary wavelength range is 100-350 nm with extension into the
visible range, but the optics are optimized at 200 nm.

The telescope general view is given in Fig.~\ref{Fig:T170M}: the
main structural elements are the Primary Mirror Unit (PMU), the
Secondary Mirror Unit (SMU) and the Instrumental Compartment (IC).
There are three attachment points of the telescope to the spacecraft
service module in the bottom frames part. The Optical Bench (OB),
carrying the scientific instruments and the Fine Guidance Sensors,
is mounted on the PMU frame. The SMU is attached to the telescope
with a spider.

\section{Focal Plane Instruments}
\label{sect:FPI}

\begin{figure}
   \vspace{2mm}
   \begin{center}
\hspace{3mm}\psfig{figure=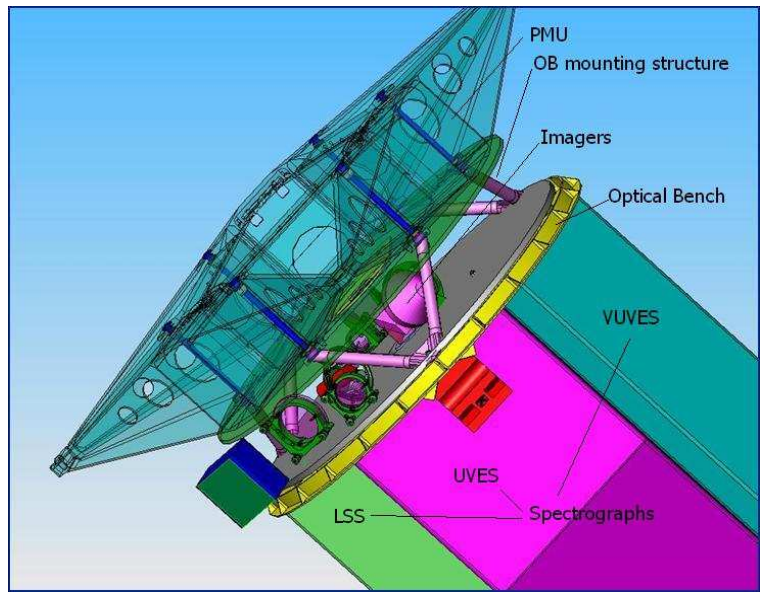,width=100mm,angle=180.0}
   \caption{Instrumental Compartment, showing the location of the focal plane instruments.   }
   \label{Fig:IC}
   \end{center}
\end{figure}

WSO-UV will have spectroscopic and imaging capabilities. The
telescope will host two spectrographs and one imager. There will be
a high resolution echelle spectrograph (\emph{High Resolution Double
Echelle Spectrograph} -- HIRDES), and a \emph{Long Slit} (low
resolution) \emph{Spectrograph} (LSS). The imager, Field Camera Unit
(FCU), will allow diffraction limited, deep UV and optical images.
Fig.~\ref{Fig:IC} shows the location of the instruments in the
Instrumental Compartment.

\subsection{HIRDES}

\begin{figure}
   \vspace{2mm}
   \begin{center}
   \hspace{3mm}\psfig{figure=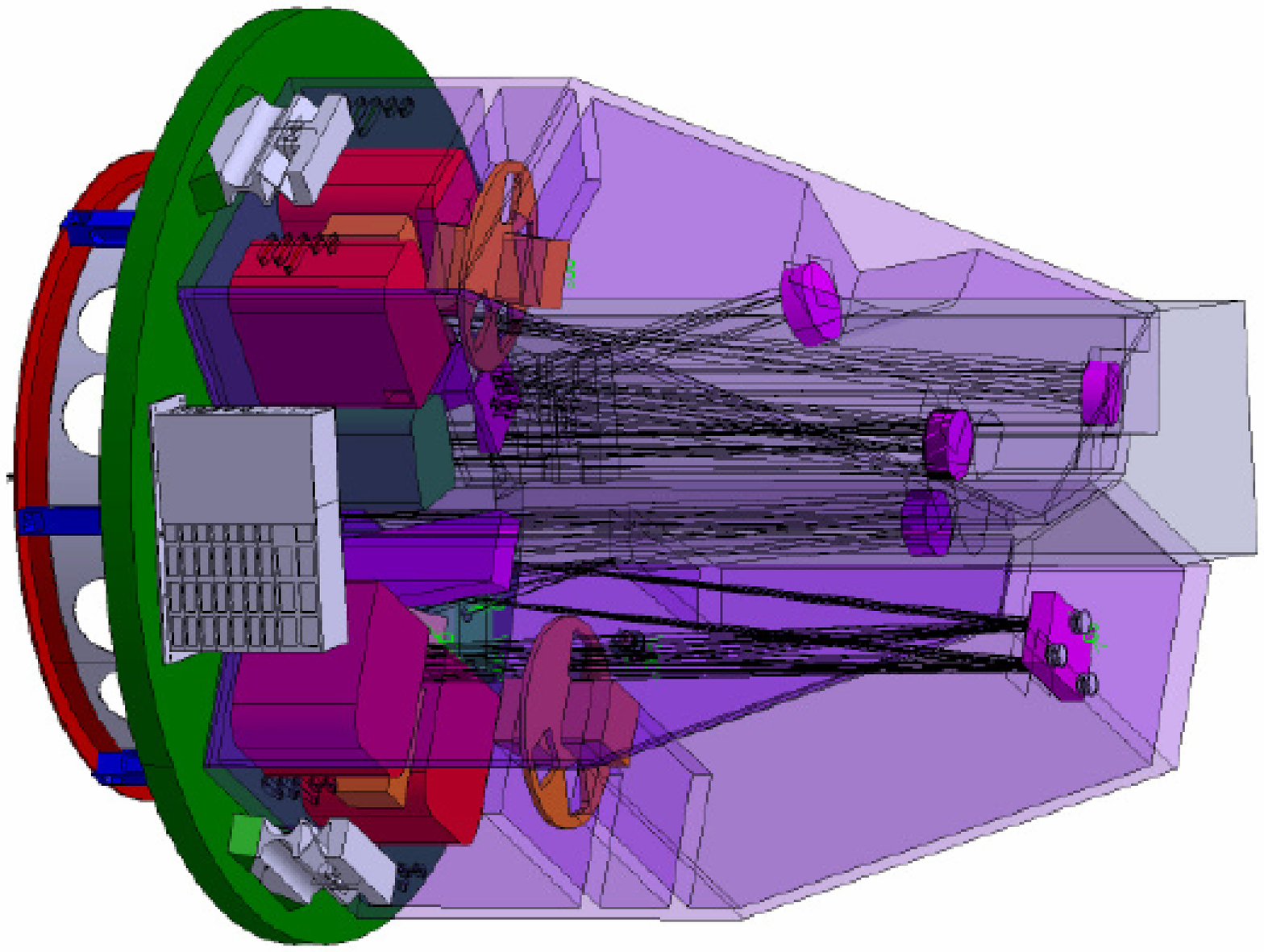,width=100mm,height=77mm,angle=270.0}
   \caption{ Schematic overview of HIRDES (Kappelmann et al. \cite{Kap06}).}
   \label{Fig:HIRDES}
   \end{center}
\end{figure}

HIRDES (Kappelmann et al. \cite{Kap06}, see Fig.~\ref{Fig:HIRDES})
design is based on the heritage of the ORFEUS (\emph{Orbiting and
Retrievable Far and Extreme Ultraviolet Spectrometer}) missions
(Barnstedt et al. \cite{Barn99}). It comprises two echelle
instruments, UVES (178-320nm) and VUVES (103-180nm), with high
spectral resolution (R$\sim$50,000). The detectors of the two
channels are photon counting devices based on Microchannel Plates,
readout by means of a Wedge\&Strip Anode based on the ORFEUS
detector design.

Each of the channels has its own entrance slit lying in the focal
surface of the T-170M telescope, on a circle with diameter 100 mm
which also host the LSS slit. The pointing will be monitored by two
(visible) sensors (one for UVES and one for VUVES) of the Internal
Fine Guidance System (IFGS), which is part of HIRDES and will allow
compensating the jitter of the spectral images due to slight
variations in the telescope pointing.

An industrial phase A study for all the spectrographs (including a
long slit low resolution one, initially designed to be accomodated
in a single box), has been completed in 2001 by Jena-Optronik
(funded by the the German Space Agency DLR). A Phase B1 study for
VUVES and UVES has been completed in May 2006 by Germany, with
Russia collaboration and with industrial support by Kaiser--Threde
(while a modified version of LSS, now considered as a separate
instrument, is now under study in China).

With the present design, the limiting UV monochromatic magnitudes
(SNR=10 in 10h) are 18 and 16 for UVES and VUVES, respectively. An
Exposure Time calculator is available on the web
(http://astro.uni-tuebingen.de/groups/wso\_uv/exptime\_calc.shtml).
The spectral resolution provided by HIRDES is similar to that
provided by HST/STIS, but higher than that which will be provided by
HST/COS. As far as sensitivity is concerned, WSO-UV/HIRDES is
comparable to HST/COS and definitely better than HST/STIS.

\subsection{LSS}

The Long Slit Spectrograph will provide low resolution
(R$\sim$1500-2500) spectra in the range 102-320 nm (possibly divided
in two sub--channels: 102-161 nm and 160-320 nm), using a
1"$\times$75" slit, with spatial resolution 0.5-1 arcsec. LSS study,
ongoing in China, is currently in phase A, with a Phase B expected
to be completed by January 2008.

\subsection{FCU}

The Field Camera Unit (FCU) will include three channels (Scuderi et
al. \cite{Scu07}):

\begin{description}
  \item[\textbf{FUV channel}]
  It covers the far UV providing medium resolution images.
To satisfy the high sensitivity requirement below 200 nm, this
channel is optimized in the range 115-190 nm, with 0.2 arcsec/pixel
scale. Since it uses the focal ratio of the telescope, this design
minimizes the numbers of optical elements (only a pick-up mirror is
required, to deviate the optical beam towards the detector),
maximizing the throughput. The FUV channel will be equipped with a
$2k\times2k$ pixels, photon counting, MCP detector (with CsI
photocathode) and will have a FoV=$6.0\times6.0$ $arcmin^{2}$. This
channel shall be equipped with a filter wheel hosting broad and
narrow band filters, neutral filters, and a
R$\sim$100 light disperser, allowing also low resolution slitless spectroscopy \\
\item[\textbf{NUV channel}] It operates in the range 150-280 nm, providing "close to diffraction limit" images at 200 nm, with a
0.03 arcsec/pixel scale.  The NUV channel will be equipped by a
$2k\times2k$ pixel MCP detector (with CsTe photocathode) and will
have a FoV=$1.0\times1.0$ $arcmin^{2}$. This channel shall be
equipped with two filter wheels hosting broad and narrow band
filters, neutral
filters, polarizers and a R$\sim$100 grism, providing slitless spectroscopic and polarimetric capabilities \\
  \item[\textbf{UVO channel}] Near ultraviolet-visual diffraction limit
imager, operating in the interval 200-1000 nm, with a 0.07
arcsec/pixel scale, equipped with a $4k\times4k$ pixels CCD,
providing a field of view of $4.6\times4.6$ $arcmin^{2}$. This
channel shall be equipped with a set of broad band and narrow band
filters, a ramp filter, polarizers and a R$\sim$250 grism.
\end{description}
\begin{figure}
   \vspace{2mm}
   \begin{center}
   \hspace{3mm}\psfig{figure=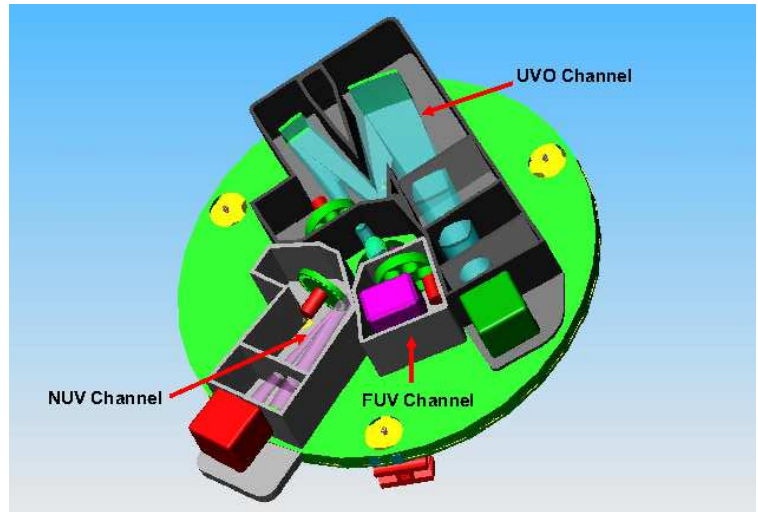,height=68mm,angle=270.0}
   \caption{Layout of the Field Camera Unit.   }
   \label{Fig:fcu}
   \end{center}
\end{figure}

\begin{figure}
   \vspace{2mm}
   \begin{center}
   \hspace{3mm}\psfig{figure=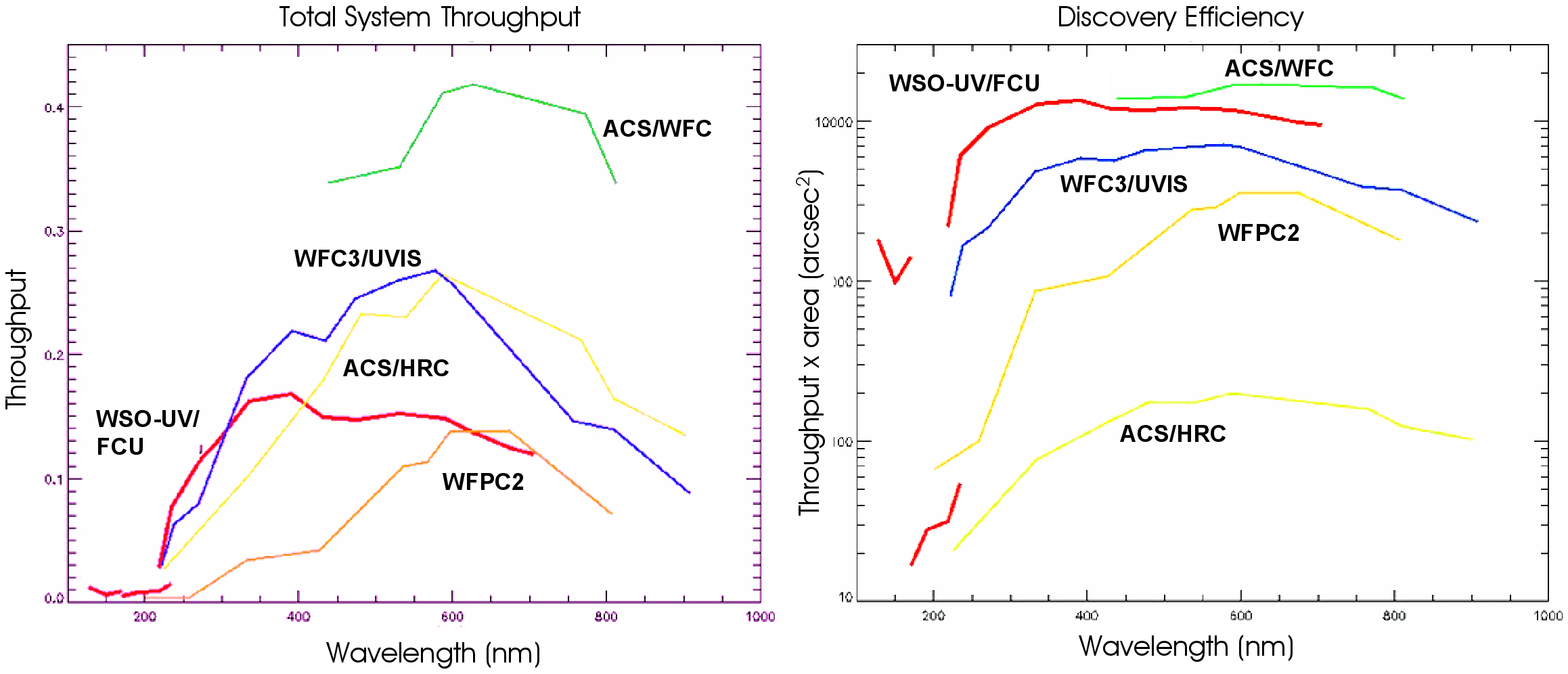,width=120mm,height=48mm,angle=270.0}
   \caption{ System throughput (a) and discovery efficiency (b) of the FCU cameras as function of the wavelength, compared with HST imagers.   }
   \label{Fig:fcuperf}
   \end{center}
\end{figure}

FUV and NUV channels, using photon counting detectors, will allow
also high time resolution observations, down to few ms time scale.
Fig.~\ref{Fig:fcuperf}a shows a preliminary estimate of the system
throughput of the three FCU channels compared to the cameras which
have flown or will flew on board of HST (WFPC2, ACS/WFC, ACS/HRC and
WFC3/UVIS, Bond H.E. et al. \cite{Bond06}). As it can be seen, the
performance of the FCU cameras compares well with the HST instrument ones.
Another useful quantity when comparing different instruments is the
discovery efficiency, plotted in Fig.~\ref{Fig:fcuperf}b, defined
as the product of the system throughput and the area of the Field of
View (FOV) as projected on the sky. Due to its large FOV, UVO has a
discovery efficiency equal or greater than ACS/WFC. In the case of
the FUV, the performance are even better when compared to HST
because no camera working in this range has a large FOV.

The Italian Space Agency founded a Phase A/B1 for the FCU. Phase A
has been completed in July 2007. Phase B1 will last four months.
\begin{acknowledgements}
The participation in the WSO-UV project in Italy is funded by Italian Space Agency under contract ASI/INAF No. I/085/06/0.
\end{acknowledgements}


\bigskip
\noindent {\bf DISCUSSION}

\bigskip
\noindent {\bf JOERN WILMS:} What is the time resolution of the UV
detectors?

\bigskip
\noindent {\bf MICHELA USLENGHI:} Time resolution will be $\sim$ 10
ms

\bigskip
\noindent {\bf JIM BEALL:} The geosynchronous orbit presents some
challenges with respect to satellite electronics. Can you comment on
this? How did you pick the orbit?

\bigskip
\noindent {\bf MICHELA USLENGHI:} We are aware of the problem and
industry is now working on the electronics design on that base. The
first choice for the orbit was an L2 one, the baseline (recently)
changed to the geosynchronous one due to cost issues.


\begin{thebibliography}{99}
\bibitem[1999]{Barn99}Barnstedt J. et al., 1999, A\&ASS, v.134, p.561-567
\bibitem[2003]{Bar03} Barstow M. A., Binette L., Brosch N.
et al., 2003, Proc. SPIE, Vol. 4854, pp. 364-374
\bibitem[2006]{Bond06} Bond H.E. et al., 2006, "Wide Field Camera 3
Instrument Mini-Handbook, Version 3.0", Baltimore STScI
\bibitem[2006]{Kap06} Kappelmann N., Barnstedt J., Gringel W. et al., 2006, Proc.
SPIE, Volume 6266, p.25
\bibitem[2007]{Pag07} Pagano I., Shustov B., Kappelmann, N., et al., 2007,  Proceedings Series of the Italian Physical Society,  F. Giovannelli \& G. Mannocchi (eds.), Vol 93, p. 691
\bibitem[2007]{Scu07} Scuderi S., Pagano I., Fiorini M., Gambicorti L., Gherardi A.,
Gianotti F., Magrin D., Miccolis M., Munari M., Pace E., Pontoni C.,
Trifoglio M., Uslenghi M., and  Shustov B., 2007, Mem.S.A.It, in
press
\end{thebibliography}
\end{document}